\documentclass[showpacs,preprintnumbers,prd,nofootinbib,floats,amssymb,floatfix]{revtex4}
\usepackage{graphicx}
\usepackage{amsxtra}
\usepackage{hyperref}
\usepackage{amssymb}
\usepackage{amstext}
\usepackage{amsmath}
\usepackage{cleveref}
\usepackage{blkarray}
\begin{document}

\title{The Hamilton-Jacobi analysis  for higher-order modified gravity}
\author{Alberto Escalante}  \email{aescalan@ifuap.buap.mx}
\author{Aldair Pantoja}  \email{jpantoja@ifuap.buap.mx}
 \affiliation{Instituto de F\'isica, Benem\'erita Universidad Aut\'onoma de Puebla. \\ Apartado Postal J-48 72570, Puebla Pue., M\'exico, }
\
 
\begin{abstract}
The Hamilton-Jacobi $[HJ]$ study for  the  Chern-Simons $[CS]$ modification of general relativity $[GR]$ is performed.  The complete structure of the Hamiltonians and the generalized brackets are reported, from these  results the $HJ$ fundamental differential is constructed and the   symmetries of the theory are found. By using the Hamiltonians   we remove an apparent  Ostrogradsky's instability and the new structure of the hamiltonian is reported. In addition,   the  counting of physical   degrees of freedom is developed and some remarks are discussed. 
\end{abstract}
 \date{\today}
\preprint{}
\maketitle
%----------------------------------------------------------------------------------------------------------------------------------------------------------------------------------------------

\section{Introduction}
It is well-known that $GR$ is a  successful framework for describing  the classical behavior of the gravitational field and its relation with the geometry of space-time \cite{Einstein1,Einstein2, Dyson, Abbott, TEHTC, Turyshev}. From the canonical point of view, $GR$ is a background independent gauge theory with  diffeomorphisms invariance; the extended Hamiltonian  is a linear combination of first class constraints  and propagates two physical  degrees of freedom \cite{DeWitt}. From the quantum point of view,  the quantization program  of gravity  is a difficult task to perform. In fact, from the nonperturbative scheme,  the non-linearity of the gravitational field, manifested in the constraints, obscures the quantization  making the complete description of a nonperturbative quantum theory of gravity still an open problem \cite{Rov, Thie}. On the other hand, the perturbative point of view of the path-integral method leads to the non-renormalizability problem \cite{Kiefer,Deser} with  all the tools that have been developed  in   quantum field theory have not worked successfully.  In this respect, it is common to study modified theories of gravity in order to obtain insights in the classical  or  quantum regime; with the expectation that these theories will provide new ideas or allow the development of new tools to carry out the quantization program, with an example of this being the so-called higher order theories \cite{Weyl, Bach, Chen, Alkac}. In fact, higher-order theories are good candidates for fixing  the infinities   that appear in the renormalization problem of quantum gravity. It is claimed  that adding higher order terms quadratic in the curvature to  gravity could help avoid this problem; since these terms have a dimensionless coupling constant, which ensures that the final theory is divergence-free \cite{Stelle, frad}.  The study of higher-order theories is a modern topic in physics, these theories are relevant in  dark energy physics \cite{Gib, Wo}, generalized electrodynamics \cite{Podolsky, Pod2, Pod3} and string theories \cite{Pol, ELi}. Furthermore, an interesting model in four dimensions can be found in the literature, in which  the Einstein-Hilbert $[EH]$ action is extended  by the  addition of a Chern-Simons four-current coupled with an auxiliary field, thus, under a particular  choice of the auxiliary field the resulting   action will be a close model to  $GR$ \cite{Jackiw}. In fact, at  Lagrangian level  the theory describes the propagation of  two degrees of freedom corresponding to  gravitational waves traveling with velocity $c$, but these propagate with different polarization intensities violating spatial reflection symmetry. Moreover, the  Schwarzchild metric is a solution of the equations of motion, thus, the modified theory  and the   $EH$ action share the same classical tests.  On the other hand, at hamiltonian level the theory is  a higher-order gauge  theory \cite{Aldair1} whose Hamiltonian analysis is known not to be easy to perform. In this respect,  the analysis of  constrained  higher-order systems is usually developed by using  the Ostrogradsky-Dirac $[OD]$ \cite{Ostrogradsky, Dirac1, Dirac2, Henneaux} or the Gitman-Lyakhovich-Tyutin $[GLT]$ \cite{Gitman1, Gitman2} methods.  $OD$ scheme is based on the extension of the phase space by considering to the fields and their velocities  as canonical coordinates and then introducing an extensión to their canonical momenta. However, the identification of the constraints is not  easy  to develop;  in some cases,   the constraints are   fixed by hand in order to obtain a consistent algebra \cite{Barcelos} and this yields the opportunity to work with alternative methods. On the other hand, the  $GLT$ framework is based on  the introduction of extra variables  which transforms  a problem with higher time derivatives to one with only first-order ones then,  by using  the Dirac brackets  the second class constraints and the extra variables can be removed \cite{Esca}.  \\
Nevertheless, there is an alternative scheme for analyzing higher-order theories: the so-called Hamilton-Jacobi method.  The $HJ$ scheme for regular  field theories  was developed by Güler  \cite{Guler1, Guler2} and later extended for  singular systems   in  \cite{Guler3, Guler4}. It is based on the identification of the constraints,  called Hamiltonians. These   Hamiltonians can
be either involutive or non-involutive and they  are used for constructing a generalized differential, where the characteristic equations,
the gauge symmetries, and the generalized $HJ$  brackets of the theory can be identified. It is important to remark that  the identification of the Hamiltonians  is performed by  means of the  null vectors, thus, the Hamiltonians  will have the  correct structure without fix them by hand as is done in other approaches, then the  identification of the symmetries will be, in general,  more economical than other schemes \cite{Bertin1, Bertin2, pimentel,  esca, esca2}. \\
With all of  above the aims of this paper is to develop a detailed $HJ$ analysis of  the  theory reported in \cite{Jackiw}. In fact, we  shall  analyze  this model beyond the Lagrangian approach reported in \cite{Jackiw}; we shall  see that the Jackiw-Yi $[JY]$ model  is a higher-order theory and it is mandatory to study this theory due to  its closeness with $GR$. However,  it is well-known that in higher-order theories could  be present ghost degrees of freedom  associated to  Ostrogradsky's instabilities \cite{plus}, namely,  the hamiltonian function is unbounded and this is reflected with the presence of    linear terms of the canonical momenta  in the hamiltonian. In this respect, it is important to comment that if  there are constraints, then  it is possible to heal those instabilities \cite{Ganz, Tai};  in our case the $JY$ model will show an apparent Ostrogradsky's instability since linear terms in the momenta will appear, however,  we will  heal the theory by using the  complete set of  Hamiltonians,  thereby exorcising the associated ghosts. \\
The paper is organized as follows. In Sect. II, we start with the $CS$ modification of $GR$,  we will work in the perturbative context, say, we will expand the metric around the Minkowski background. We shall observe that the modified theory is of higher-order in the temporal derivatives, then we shall introduce a change of variables in order to express the action in terms of only  first-order temporal derivatives. The   change of variable will allows us to develop the  $HJ$ analysis in an easy way;  the identification of the Hamiltonians, the construction of the generalized differential and the symmetries will be identified directly.  In Sect. III we present the conclusions and some remarks.

\section{The Hamilton-Jacobi analysis}

The modified $EH$ action is given by \cite{Jackiw}
\begin{equation}
\label{S}
S[g_{\mu\nu}] = \int_{M}\left( R\sqrt{-g} + \frac{1}{4}\theta^{*}R^{\sigma}{}_{\tau}{}^{\mu\nu}R^{\tau}{}_{\sigma\mu\nu} \right)d^{4}x,
\end{equation}
where $M$ is the space-time manifold, $g_{\mu\nu}$ the metric tensor, $R$ the scalar curvature, $g$ the determinant of the metric, $R^{\alpha}{}_{\beta\mu\nu}$ the Riemman tensor and $\theta$ is a  coupling field. In general,  $\theta$  can be viewed as an external quantity or as a local dynamical variable, however, in order to obtain an action close  to $GR$ we  are going to choose  $\theta=\frac{t}{\Omega}$. Along the paper we will use grek letters for labeling space-time indices $\mu=0,1,2,3$ and latin letters for space indices $i=1,2,3$. In addition, we will work within the perturbative context  expanding  the metric around the Minkowski background
\begin{equation}
\label{g=eta+h}
g_{\mu\nu} = \eta_{\mu\nu} + h_{\mu\nu},
\end{equation}
where $h_{\mu\nu}$ is the perturbation. By substituting the expression for $\theta$ and by taking into account eq. \eqref{g=eta+h} in \eqref{S} we obtain the following linearized action
\begin{equation}
\label{S[h]}
S[h_{\mu\nu}]=-\frac{1}{2}\int_{M} h^{\mu\nu}\left(G_{\mu\nu}^{lin}+C_{\mu\nu}^{lin}\right)d^{4}x,
\end{equation}
where $G_{\mu\nu}^{lin}$ is the linearized version of the Einstein tensor and $C_{\mu\nu}^{lin}$ is a linearized Cotton-type tensor $C_{\mu\nu}^{lin}=-\frac{1}{4\Omega}[\epsilon_{0\mu\lambda\gamma}\partial^{\lambda}(\square h^{\gamma}{}_{\nu}-\partial_{\nu}\partial_{\alpha}h^{\alpha\gamma})+\epsilon_{0\nu\lambda\gamma}\partial^{\lambda}(\square h^{\gamma}{}_{\mu}-\partial_{\mu}\partial_{\alpha}h^{\alpha\gamma})]$ \cite{Jackiw} defined in four-dimensions.  
\\
Now we shall suppose that the space-time has a topology  $M\cong\mathbb{R}\times\Sigma$, where  $\mathbb{R}$ is an evolution parameter and $\Sigma$ is a Cauchy hypersurface. Hence,  by performing the  $3+1$ decomposition of the action \eqref{S[h]} we write down the corresponding Lagrangian density
\begin{eqnarray}
\label{Lag4}
\nonumber
\mathcal{L}&=&\int\left[\frac{1}{2}\dot{h}_{ij}\dot{h}^{ij}-\partial_{j}h_{0i}\partial^{j}h^{0i}-\frac{1}{2}\partial_{k}h_{ij}\partial^{k}h^{ij}-\frac{1}{2}\dot{h}^{i}{}_{i}\dot{h}^{j}{}_{j}+\partial^{j}h^{0}{}_{0}\partial_{j}h^{i}{}_{i}+\frac{1}{2}\partial_{k}h^{i}{}_{i}\partial^{k}h^{j}{}_{j}-2\partial^{i}h^{0}{}_{i}\dot{h}^{j}{}_{j}\right.
\\
\nonumber
&& -\partial_{i}h^{0}{}_{0}\partial_{j}h^{ij}-\partial_{i}h^{ij}\partial_{j}h^{k}{}_{k}+2\partial_{j}h^{0}{}_{i}\dot{h}^{ij}+\partial_{i}h^{i}{}_{0}\partial_{j}h^{0j}+\partial_{k}h^{k}{}_{i}\partial_{j}h^{ij}+\frac{1}{\mu}\epsilon^{ijk}(-\ddot{h}^{l}{}_{i}\partial_{j}h_{lk}
\\
\label{L}
&& \left. +2\dot{h}^{l}{}_{i}\partial_{j}\partial_{l}h^{0}{}_{k}+\partial_{l}h^{m}{}_{i}\partial_{m}\partial_{j}h^{l}{}_{k}+\nabla^{2}h^{0}{}_{i}\partial_{j}h_{0 k}+\nabla^{2}h^{m}{}_{i}\partial_{j}h_{m k})\right]d^{3}x,
\end{eqnarray}
where we have  defined $\mu\equiv2\Omega$ and $\epsilon^{ijk}\equiv\epsilon_{0}{}^{ijk}$. As it was commented  above, we will   reduce the order of the time derivatives of the  Lagrangian (\ref{Lag4}) by extending the configuration space, this is done by introducing the following change of variable 
\begin{equation}
\label{K}
K_{ij}  = \frac{1}{2}(\dot{h}_{ij}-\partial_{i}h_{0j}-\partial_{j}h_{0i}),
\end{equation}
here $K_{ij}$ is related with the so-called extrinsic curvature \cite{Frankel, Furi}. Thus,  by substituting \eqref{K} into \eqref{L} we rewrite  the Lagrangian in the following  new fashion 
\begin{eqnarray}
\label{Lag}
\nonumber 
\mathcal{L} &=& \int \left[ 2K_{ij}K^{ij} - 2K^{i}{}_{i}K^{j}{}_{j} - h_{00}R_{ij}{}^{ij} - h_{ij}R^{ij} + \frac{1}{2}h^{i}{}_{i}R_{ij}{}^{ij} + \frac{1}{\mu}\epsilon^{ijk}( 4K_{i}{}^{l}\partial_{j}K_{kl} + \partial^{m}h_{im}\partial_{j}\partial^{l}h_{kl} \right.
\\
&& \left. + \nabla^{2}h_{i}{}^{m}\partial_{j}h_{km}) + \psi^{ij}(\dot{h}_{ij}-\partial_{i}h_{0j}-\partial_{j}h_{0i}-2K_{ij}) \right]d^{3}x,
\end{eqnarray}
where we have added the  Lagrange multipliers $\psi^{ij}$  enforcing the the relation (\ref{K}),  and the expressions  $R_{ij}{}^{ij}$ and $R_{ij}$ are defined in the following way 
\begin{eqnarray}
R_{ij}{}^{ij} &\equiv& \partial^{i}\partial^{j}h_{ij} - \nabla^{2}h^{i}{}_{i},
\\
R_{ij} &\equiv& \frac{1}{2}(\partial_{i}\partial^{k}h_{jk}+\partial_{j}\partial^{k}h_{ik}-\partial^{i}\partial^{j}h^{k}{}_{k}-\nabla^{2}h_{ij}).
\end{eqnarray}
Now, we calculate   the  canonical momenta associated with the dynamical variables 
\begin{eqnarray}
\label{pi00}
\pi^{00} &=& \frac{\partial\mathcal{L}}{\partial\dot{h}_{00}} = 0,
\\
\nonumber
\\
\label{pi0i}
\pi^{0i} &=& \frac{\partial\mathcal{L}}{\partial\dot{h}_{0i}} = 0,
\\
\nonumber
\\
\label{piij}
\pi^{ij} &=& \frac{\partial\mathcal{L}}{\partial\dot{h}_{ij}} = \psi^{ij},
\\
\nonumber
\\
\label{Pij}
P^{ij} &=& \frac{\partial\mathcal{L}}{\partial\dot{K}_{ij}} = 0,
\\
\nonumber
\\
\label{Gammaij}
\Lambda^{ij} &=& \frac{\partial\mathcal{L}}{\partial\dot{\psi}_{ij}} = 0.
\end{eqnarray}
Thus,  from the  equations \eqref{pi00}-\eqref{Gammaij} we identify the following $HJ$ Hamiltonians of the theory 
\begin{eqnarray}
\label{H_{0}}
\mathcal{H}' &\equiv& \mathcal{H}_{0} + \Pi = 0,
\\
\label{H_{1}}
H_{1}^{00} &\equiv& \pi^{00} = 0,
\\
\label{H_{2}}
H_{2}^{0i} &\equiv& \pi^{0i} = 0,
\\
\label{H_{3}}
H_{3}^{ij} &\equiv& \pi^{ij} - \psi^{ij} = 0,
\\
\label{H_{4}}
H_{4}^{ij} &\equiv& P^{ij} = 0,
\\
\label{H_{5}}
H_{5}^{ij} &\equiv& \Lambda^{ij} = 0,
\end{eqnarray}
where $\mathcal{H}_{0}$ is the canonical hamiltonian defined as usual  $\mathcal{H}_{0}=\dot{h}_{\mu\nu}\pi^{\mu\nu}+\dot{K}_{ij}P^{ij}+ \dot{\psi}_{ij}\Lambda^{ij}-\mathcal{L}$ and $\Pi=\partial_{0}S$ \cite{Bertin1, Bertin2, pimentel, esca, esca2}. Moreover,  the fundamental Poisson brackets $[PB]$ between the canonical variables are given by 
\begin{eqnarray}
\lbrace h_{\mu\nu},\pi^{\alpha\beta} \rbrace &=& \frac{1}{2}(\delta_{\mu}^{\alpha}\delta_{\nu}^{\beta}+\delta_{\nu}^{\alpha}\delta_{\mu}^{\beta})\delta^{3}(x-y),
\\
\lbrace K_{ij},\pi^{kl} \rbrace &=& \frac{1}{2}(\delta_{i}^{k}\delta_{j}^{l}+\delta_{j}^{k}\delta_{i}^{l})\delta^{3}(x-y),
\\
\label{FPB}
\lbrace \psi^{ij},\Lambda_{kl} \rbrace &=& \frac{1}{2}(\delta^{i}_{k}\delta^{j}_{l}+\delta^{j}_{k}\delta^{i}_{l})\delta^{3}(x-y).
\end{eqnarray}
Furthermore, in the $HJ$ scheme, the dynamics of the system is governed by the fundamental differential defined as 
\begin{equation}
\label{d}
dF=\lbrace F ,H_{I}\rbrace d\omega^{I},
\end{equation}
where $F$ is any function defined on the phase space, $H_{I}$ is the set of all Hamiltonians \eqref{H_{0}}-\eqref{H_{5}} and $\omega^{I}$ are the parameters related to them. It is important to remark,  that in the $HJ$ method the Hamiltonians are classified as  involutive and non-involutive. Involutive ones   are those whose $PB$ with all Hamiltonians, including themselves, vanish; otherwise, they are called non-involutive. Because of integrability conditions, the non-involutive Hamiltonians are removed from the fundamental differential (\ref{d})  by introducing the so-called generalized  brackets, these new brackets are given by 
\begin{equation}
\label{DB}
\lbrace f,g \rbrace^{*} = \lbrace f,g \rbrace - \lbrace f,H^{a'} \rbrace C_{a'b'}^{-1} \lbrace H^{b'},g \rbrace,
\end{equation}
where $C^{a'b'}$ is the matrix  formed with the $PB$ between all non-involutive Hamiltonians. From   \eqref{H_{0}}-\eqref{H_{5}}  the non-involutive Hamiltonians are  $H_{3}^{ij}$ and $H_{5}^{ij}$, whose $PB$ is 
\begin{eqnarray}
\lbrace H_{3}^{ij},H_{5}^{ij} \rbrace = - \frac{1}{2}(\eta^{ik}\eta^{jl}+\eta^{il}\eta^{kj})\delta^{3}(x-y),
\end{eqnarray}
therefore, the matrix $C^{a'b'}$ given by  
\begin{equation}
C^{a'b'}
=
\bordermatrix{
 & &  \cr
& 0 & - \frac{1}{2}(\eta^{ik}\eta^{jl}+\eta^{il}\eta^{kj}) \cr
& \frac{1}{2}(\eta^{ik}\eta^{jl}+\eta^{il}\eta^{kj}) & 0 \cr
}
\delta^{3}(x-y),
\end{equation}
and  its inverse $C_{a'b'}^{-1}$ takes the  form
\begin{equation}
C_{a'b'}^{-1}
=
\begin{pmatrix}
0 & \frac{1}{2}(\eta^{ik}\eta^{jl}+\eta^{il}\eta^{kj}) \cr
- \frac{1}{2}(\eta^{ik}\eta^{jl}+\eta^{il}\eta^{kj}) & 0 \cr
\end{pmatrix}
\delta^{3}(x-y).
\end{equation}
In this manner, the following non-vanishing  generalized brackets between the fields arise 
\begin{eqnarray}
\label{PB1}
\lbrace h_{\mu\nu},\pi^{\alpha\beta} \rbrace^{*} &=&  \frac{1}{2}(\delta_{\mu}^{\alpha}\delta_{\nu}^{\beta}+\delta_{\mu}^{\beta}\delta_{\nu}^{\alpha})\delta^{3}(x-y),
\\
\label{PB2}
\lbrace K_{ij},P^{kl} \rbrace^{*} &=&  \frac{1}{2}(\delta_{i}^{k}\delta_{j}^{l}+\delta_{i}^{l}\delta_{j}^{k})\delta^{3}(x-y),
\\
\label{PB3}
\lbrace h_{\mu\nu},\psi^{\alpha\beta} \rbrace^{*} &=&  \frac{1}{2}(\delta_{\mu}^{\alpha}\delta_{\nu}^{\beta}+\delta_{\mu}^{\beta}\delta_{\nu}^{\alpha})\delta^{3}(x-y), 
\\
\label{PB4}
\lbrace \psi_{ij},\Lambda^{kl} \rbrace^{*} &=& 0,
\end{eqnarray}
we observe from \eqref{PB4} that  the canonical variables $(\psi_{ij},\Lambda^{kl})$ can be removed which  implies that we can perform  the substitution of  $\pi^{ij}=\psi^{ij}$ and $\Lambda^{ij}=0$, hence, the canonical   hamiltonian takes the form
\begin{eqnarray}
\nonumber
\mathcal{H}_{0} &=& \int[ 2K^{i}{}_{i}K^{j}{}_{j} - 2K_{ij}K^{ij} + h_{00}R_{ij}{}^{ij} + h_{ij}R^{ij} - \frac{1}{2}h^{i}{}_{i}R_{ij}{}^{ij} - \frac{1}{\mu}\epsilon^{ijk}( 4K_{i}{}^{l}\partial_{j}K_{kl} + \partial^{m}h_{im}\partial_{j}\partial^{l}h_{kl}
\\
\label{CanonicalHamiltonian}
&& + \nabla^{2}h_{i}{}^{m}\partial_{j}h_{km}) - 2h_{0j}\partial_{i}\pi^{ij} + 2K_{ij}\pi^{ij} ]d^{3}x.
\end{eqnarray}
It is worth to comment, that the canonical hamiltonian  has  linear terms in the  momenta $\pi^{ij}$ and this  fact could be related to Ostrogradsky's instabilities. Nevertheless,  it is well-known that those  instabilities could  be  healed   by means the correct identification of the constraints \cite{Ganz, Tai}. In this respect, an advantage of the $HJ$ scheme is that the constraints are identified directly and it is not necessary to fix them by hand, then  with the generalized brackets and the identification of the Hamiltonians  we can remove the linear canonical momenta terms. In fact, by using the Hamiltonians \eqref{H_{0}}-\eqref{H_{5}} the canonical hamiltonian takes the following form   
\begin{eqnarray}
\nonumber
\mathcal{H}_{0}' &=& \int[ \frac{1}{2}\pi^{ij}\pi_{ij} - \frac{1}{4}\pi^{i}{}_{i}\pi^{j}{}_{j} + h_{ij}R^{ij} - \frac{1}{\mu}\epsilon^{ijk}(4K_{i}{}^{l}\partial_{j}K_{kl} + \partial^{m}h_{im}\partial_{j}\partial^{l}h_{kl} + \nabla^{2}h_{i}{}^{l}\partial_{j}h_{kl}) 
\\
\nonumber
&& - \frac{4}{\mu^{2}}(2\partial^{i}K_{ij}\partial^{j}K^{k}{}_{k}+2\partial^{i}K^{j}{}_{k}\partial_{i}K_{j}{}^{k}-2\partial^{j}K^{i}{}_{k}\partial_{i}K_{j}{}^{k}-\partial^{j}K^{i}{}_{k}\partial^{k}K_{ij}-\partial_{k}K^{i}{}_{i}\partial^{k}K^{j}{}_{j}]d^{3}x.
\end{eqnarray}
 hence, the  Ostrogradsky instability has been healed and the associated ghost was  exorcised. \\ 
 On the other hand, with all these results  we rewrite the fundamental differential in terms of either involutive Hamiltonians or  generalized brackets, this is 
\begin{eqnarray}
\label{FD1}
dF &=& \int[ \lbrace F,H' \rbrace^{*} dt + \lbrace F,H_{1}^{00}\rbrace^{*} d\omega_{00}^{1} + \lbrace F,H_{2}^{0i} \rbrace^{*} d\omega_{0i}^{2} + \lbrace F,H_{4}^{ij} \rbrace^{*} d\omega_{ij}^{4} ]d^{3}y.
\end{eqnarray}
 thus,  we will search if there are more Hamiltonians in the theory. For this aim,  we shall take  into account  either the generalized differential  \eqref{FD1} or   the Frobenius integrability conditions which,  ensure that system is integrable, this is  
\begin{equation}
\label{IC}
dH_{a}=0,
\end{equation}
where $H_a \equiv(H_1^{00}, H_2^{0i}, H_4^{ij})$ are all involutive Hamiltonians. From  integrability conditions \eqref{IC}  the following 10 new Hamiltonians arise 
\begin{eqnarray}
\label{H_{6}}
H_{6}^{00} &\equiv& \nabla^{2}h^{i}{}_{i} - \partial^{i}\partial^{j}h_{ij} = 0,
\\
\label{H_{7}}
H_{7}^{0i} &\equiv& \partial_{j}\pi^{ij} = 0,
\\
\label{H_{8}}
H_{8}^{ij} &\equiv& \pi^{ij} - 2K^{ij} + 2\eta^{ij}K^{k}{}_{k} - \frac{2}{\mu}(\epsilon^{ikl}\eta^{jm} + \epsilon^{jkl}\eta^{im})\partial_{k}K_{lm} = 0,
\end{eqnarray}
Now, we observe that the Hamiltonians $H_4^{ij}$, $H_6^{00}$ and $H_8$ are non-involutive, therefore they will be  removed by introducing a new set of generalized brackets. In this respect, if we calculate the matrix whose entries will be  all generalized brackets, say (\ref{PB1})-(\ref{PB4}), between the non-involutive Hamiltonians, we will find null vectors, say  $v^i=( \frac{1}{2} \partial_i \partial_j\zeta, \delta^i{_{k}} \zeta,0)$, where $\zeta$ is an arbitrary function. Hence,  from the contraction of the null vectors with the Hamiltonians \cite{esca, esca2}, we will find the following involutive Hamiltonian
\begin{eqnarray}
H_{9} &=& \nabla^{2}h^{i}{}_{i} - \partial^{i}\partial^{j}h_{ij} + \frac{1}{2}\partial_{i}\partial_{j}P^{ij},
\end{eqnarray}
thus,  there are only 12 non-involutive Hamiltonians  $(H_{4}^{ij},H_{8}^{ij})$ whose generalized  brackets are  given by 
\begin{eqnarray}
\nonumber
\lbrace H_{4}^{ij},H_{8}^{ij} \rbrace^{*} &=& 2[ \frac{1}{2\mu}(\epsilon^{ikm}\eta^{jl}+\epsilon^{jkm}\eta^{il}+\epsilon^{ilm}\eta^{jk}+\epsilon^{ilm}\eta^{ik})\partial_{m} + \frac{1}{2}(\eta^{ik}\eta^{jl}
\\
\label{(H4,H8)}
&& +\eta^{jk}\eta^{il}) - \eta^{ij}\eta^{kl} ]\delta^{3}(x-y).
\end{eqnarray}
In this manner, we proceed to construct the new set of $HJ$ generalized brackets, namely $\lbrace\;,\;\rbrace^{**}$,  in the same way as we did before  with the brackets \eqref{PB1}-\eqref{PB4}. The non-trivial new generalized brackets are given by 
\begin{eqnarray}
\label{GB1}
\lbrace h_{ij},\pi^{kl} \rbrace^{**} &=& \frac{1}{2}(\delta_{i}^{k}\delta_{j}^{l}+\delta_{i}^{l}\delta_{j}^{k})\delta^{3}(x-y),
\\
\label{GB2}
\lbrace K_{ij},P^{kl} \rbrace^{**} &=& 0,
\\
\nonumber
\lbrace h_{ij},K_{kl} \rbrace^{**} &=& \frac{1}{4}(\eta_{ik}\eta_{jl}+\eta_{il}\eta_{jk}-\eta_{ij}\eta_{kl})\delta^{3}(x-y) + \frac{\mu^{2}}{4\Xi}[[(\eta_{ik}\eta_{jl}+\eta_{il}\eta_{jk}-\eta_{ij}\eta_{kl})\nabla^{2} + (\eta_{ij}\partial_{k}\partial_{l}
\\
\nonumber
&& +\eta_{kl}\partial_{i}\partial_{j})](\nabla^{2}+\mu^{2}) - 3\partial_{i}\partial_{j}\partial_{k}\partial_{l} - \frac{3\mu^{2}}{4}(\eta_{ik}\partial_{j}\partial_{l}+\eta_{il}\partial_{j}\partial_{k} +\eta_{jk}\partial_{i}\partial_{l}+\eta_{jl}\partial_{i}\partial_{k})
\\
\nonumber
&& + \frac{\mu}{4}[(\epsilon_{ik}{}^{m}\eta_{jl}+\epsilon_{jk}{}^{m}\eta_{il}+\epsilon_{il}{}^{m}\eta_{jk}+\epsilon_{jl}{}^{m}\eta_{ik})(\nabla^{2}+\mu^{2}) + 3(\epsilon_{ik}{}^{m}\partial_{j}\partial_{l}+\epsilon_{jk}{}^{m}\partial_{i}\partial_{l}
\\
\label{GB3}
&& +\epsilon_{il}{}^{m}\partial_{j}\partial_{k}+\epsilon_{jl}{}^{m}\partial_{i}\partial_{k})]\partial_{m}]\delta^{3}(x-y),
\end{eqnarray}
where $\Xi\equiv-\mu^{2}(\nabla^{2}+\mu^{2})(\nabla^{2}+\frac{\mu^{2}}{4})$. It is worth commenting, that  some   brackets were reported in \cite{Aldair1}, however, there are some differences. In fact, in this paper we have used an alternative analysis and new variables were introduced; the introduction of the variables allowed us to identify  the brackets (\ref{GB3})  directly and they have a more  compact form than those  reported in \cite{Aldair1}. Moreover, the tedious classification of the constrains into first class and  second class as usually is done, in the $HJ$ scheme it is  not necessary.  Thus, we can observe that the $HJ$ is more economical. \\
With the new set of either involutives Hamiltonians or generalized brackets, the  fundamental differential takes the following new form 
\begin{eqnarray}
\nonumber
dF &=& \int[ \lbrace F,H' (y)\rbrace^{**}dt + \lbrace F,H_{1}^{00}(y) \rbrace^{**}d\omega_{00}^{1} + \lbrace F,H_{2}^{0i} (y)\rbrace^{**}d\omega_{0i}^{2} + \lbrace F,H_{7}^{0i} (y)\rbrace^{**}d\omega_{0i}^{7}  \nonumber \\ &+& \lbrace F,H_{9} (y)\rbrace^{**}d\omega^{9} ]d^{3}y,
\label{FD2}
\end{eqnarray}
where
\begin{eqnarray}
H_{1}^{00} &=& \pi^{00},
\\
H_{2}^{0i} &=& \pi^{0i},
\\
H_{7}^{0i} &=& \partial_{j}\pi^{ij},
\\
H_{9} &=& \nabla^{2}h^{i}{}_{i} - \partial^{i}\partial^{j}h_{ij}.
\label{ham}
\end{eqnarray}
From integrability conditions of $ H_7^{0i}$ and $H_9$  we find 
\begin{eqnarray}
\label{dH7}
dH_{7}^{0i} &=& 0,
\\
\label{dH8}
dH_{9} &=& - \partial_{i}\partial_{j}\pi^{ij} = -\partial_{i} H_{7}^{0i} = 0,
\end{eqnarray}
therefore, there are not  further  Hamiltonians. It is worth to comment,  that the Hamiltonians given in (\ref{ham}) are related to those reported in \cite{Bertin} where only linearized gravity was studied. However, there are differences: from on  side, the PB reported in \cite{Bertin} and the generalized brackets found  in (\ref{GB1})-(\ref{GB3}) are different. On the other hand, the contribution of the modification is present in the generalized brackets,  and this fact  will be relevant in the study of quantization  because the generalized brackets will be changed to commutators and the contribution could  provide differences with respect standard linearized gravity.   \\
Now, we will  calculate the $HJ$ characteristic equations, they are given by 
\begin{eqnarray}
\label{CE1}
dh_{00} &=& d\theta_{00}^{1},
\\
\label{CE2}
dh_{0i} &=& \frac{1}{2}d\theta_{0i}^{2},
\\
\label{CE3}
dh_{ij} &=& [2K_{ij}+\partial_{i}h_{0j}+\partial_{j}h_{0i}]dt - \frac{1}{2}(\delta_{i}^{k}\partial_{j}+\delta_{j}^{k}\partial_{i})d\theta_{0k}^{7},
\\
\label{CE4}
d\pi^{00} &=& - R_{ij}{}^{ij}dt,
\\
\label{CE5}
d\pi^{0i} &=& \frac{1}{2}\partial_{j}\pi^{ij}dt,
\\
\nonumber
d\pi^{ij} &=& [\eta^{ij}\nabla^{2}h_{00} - \partial^{i}\partial^{j}h_{00} - \eta^{ij}R_{kl}{}^{kl} - 2R^{ij} - \frac{1}{\mu}[(\epsilon^{ikl}\partial^{j}+\epsilon^{jkl}\partial^{i})\partial_{k}\partial^{m}h_{lm}
\\
\label{CE6}
&& - (\epsilon^{ikl}\eta^{jm}+\epsilon^{jkl}\eta^{im})\partial_{k}\nabla^{2}h_{lm}]]dt + (\partial^{i}\partial^{j} - \eta^{ij}\nabla^{2})d\theta^{9},
\\
\label{CE7}
dK_{ij} &=& [ - \frac{1}{2}\partial_{i}\partial_{j}h_{00} - R_{ij} + \frac{1}{4}\eta_{ij}R_{kl}{}^{kl} ]dt + \frac{1}{2}\partial_{i}\partial_{j}d\theta_{9},
\\
\label{CE8}
dP^{ij} &=& [0]dt,
\end{eqnarray}
from the characteristic equations we can identify the following facts: from   equations \eqref{CE1}-\eqref{CE2} we observe  that the variables $h_{00}$ and $h_{0i}$ are identified as Lagrange multipliers. Moreover,  from \eqref{GB2} and \eqref{CE8} we discard to  $P^{ij}$ as degree of freedom because its time  evolution vanishes. Furthermore, we identify  the equations of motion for $h_{ij}$ and its momentum $\pi^{ij}$. In fact,  by taking  $d\theta_{0k}^{7}=0$ and $d\theta^{9}=0$,  we obtain
\begin{eqnarray}
\label{eqh}
\dot{h}_{ij} &=& 2K_{ij} + \partial_{i}h_{0j}+\partial_{j}h_{0i},
\\
\nonumber
\dot{\pi}^{ij} &=& \eta^{ij}\nabla^{2}h_{00} - \partial^{i}\partial^{j}h_{00} - \eta^{ij}R_{kl}{}^{kl} - 2R^{ij} - \frac{1}{\mu}[(\epsilon^{ikl}\partial^{j}+\epsilon^{jkl}\partial^{i})\partial_{k}\partial^{m}h_{lm}
\\
&& - (\epsilon^{ikl}\eta^{jm}+\epsilon^{jkl}\eta^{im})\partial_{k}\nabla^{2}h_{lm}], \\ 
\dot{K}_{ij} &=& - \frac{1}{2}\partial_{i}\partial_{j}h_{00} - R_{ij} + \frac{1}{4}\eta_{ij}R_{kl}{}^{kl}.
\end{eqnarray}
 We observe that (\ref{eqh}) corresponds to the definition of $K_{ij}$, thus, if we use $(\ref{eqh})$ and $\dot{K}_{ij}$ we will obtain a second order time equation for $h_{ij}$ as expected, then there are six degrees of freedom associated with the perturbation. In this manner,  we calculate the number of physical degrees of freedom as follows: there are 12 canonical  variables $(h_{ij}, \pi^{ij})$  and eight involutive Hamiltonians $(H_1^{00}, H_2^{0i}, H_7^{0i}, H_9)$, thus  
\begin{eqnarray}
\nonumber
DOF&=& \frac{1}{2}[12 - 8]=2,
\end{eqnarray}
and thus,  the theory has two physical degrees of freedom  just like $GR$ \cite{Jackiw, Aldair1}.  \\
On the other hand, if in the characteristics equations we take $dt=0$, then   we identify the following  canonical transformations 
\begin{eqnarray}
\label{deltah00}
\delta h_{00} &=& \delta\omega_{00}^{1},
\\
\label{deltah0i}
\delta h_{0i} &=& \frac{1}{2}\delta\omega_{0i}^{2},
\\
\label{deltahij}
\delta h_{ij} &=& - \frac{1}{2}(\delta_{i}^{k}\partial_{j} + \delta_{j}^{k}\partial_{i})\delta\omega_{0k}^{7},
\end{eqnarray}
moreover, we can then identify  the corresponding gauge transformations of the theory by considering  that the Lagrangian (\ref{Lag}) will be invariant under (\ref{deltah00})-(\ref{deltahij})  if the variation   $\delta \mathcal{S}=0$ \cite{Ber}, this is  
\begin{eqnarray}
\delta S &=& \left[ \frac{\partial S}{\partial h_{\mu\nu}}\delta h_{\mu\nu} + \frac{\partial S}{\partial(\partial_{\alpha}h_{\mu\nu})}\delta(\partial_{\alpha}h_{\mu\nu}) + \frac{\partial S}{\partial(\partial_{\alpha}\partial_{\beta}h_{\mu\nu})}\delta(\partial_{\alpha}\partial_{\beta}h_{\mu\nu}) \right]
\\
\nonumber
&=& \int \left[\left( - \square h^{\mu\nu} + \square h^{\lambda}{}_{\lambda}\eta^{\mu\nu} - \partial_{\alpha}\partial_{\lambda}h^{\alpha\lambda}\eta^{\mu\nu} - \partial^{\mu}\partial^{\nu}h^{\lambda}{}_{\lambda} + 2\partial^{\mu}\partial_{\lambda}h^{\nu\lambda} + \frac{1}{\mu}\epsilon^{0\mu\lambda\gamma}(\partial^{\nu}\partial_{\alpha}\partial_{\lambda}h^{\alpha}{}_{\gamma} \right.\right.
\\
\label{VariationL}
&& \left.\left. - \partial_{\lambda}\square h^{\nu}{}_{\gamma}) \right)\delta h_{\mu\nu} \right] d^{4}x=0, 
\end{eqnarray}
thus,  by taking  account  \eqref{deltah00}-\eqref{deltahij} into the variation,  we obtain the following 
\begin{eqnarray}
\nonumber
\delta S &=& \int[ R_{ij}{}^{ij}\delta\omega^{1}_{00} + \frac{1}{2}[2\nabla^{2}h_{0}{}^{i} + 2\partial^{i}\dot{h}^{j}{}_{j} - 2\partial^{i}\partial^{j}h_{0j} - 2\partial_{j}\dot{h}^{ij} + \frac{1}{\mu}\epsilon^{0ijk}(\partial_{j}\nabla^{2}h_{0k}-\partial_{j}\partial^{l}\dot{h}_{kl})]\delta\omega^{2}_{0i}
\\
\nonumber
&& - \frac{1}{2}[\ddot{h}^{ij} - \ddot{h}^{k}{}_{k}\eta^{ij} + 2\partial^{k}\dot{h}_{0k}\eta^{ij} - 2\partial^{i}\dot{h}_{0}{}^{j} + \partial^{i}\partial^{j}h_{00} - \nabla^{2}h_{00}\eta^{ij} + 2R^{ij} - R_{kl}{}^{kl}\eta^{ij}
\\
&& + \frac{1}{\mu}\epsilon^{0ikl}(\partial_{k}\ddot{h}^{j}{}_{l}-\partial^{j}\partial_{k}\dot{h}_{0l}+\partial^{j}\partial_{k}\partial^{m}h_{lm}-\partial_{k}\nabla^{2}h^{j}{}_{l})]\delta(\partial_{i}\omega^{7}_{0j}+\partial_{j}\omega^{7}_{0i}) ]d^{4}x=0.
\end{eqnarray}
Now,  we  define $\partial_{0}\xi\equiv\delta\omega_{00}^{1}$, so  after  long algebraic work we find that the variation takes the form  
\begin{eqnarray}
\nonumber
\delta S &=& \int [ - \partial_{j}\dot{h}^{ij} + \partial^{i}h^{j}{}_{j} + \nabla^{2}h_{0}{}^{i} - \partial^{i}\partial^{j}h_{0j} + \frac{1}{2\mu}\epsilon^{0ijk}(\partial_{j}\nabla^{2}h_{0k}-\partial_{j}\partial^{l}\dot{h}_{kl}) ] ( - \partial_{i}\xi + \delta\omega^{2}_{0i} + \partial_{0}\delta \omega^{7}_{0i}) d^{4}x,\\
&=&0, 
\end{eqnarray}
hence,  the action will be invariant under \eqref{deltah00}-\eqref{deltahij} if the  the parameters $\omega's$  obey 
\begin{equation}
\label{RelationParameters1}
\delta\omega_{0i}^{2} = - \partial_{0}\delta\omega_{0i}^{7} + \partial_{i}\xi.
\end{equation}
Now, we  will write \eqref{RelationParameters1} in a new fashion. In fact, we introduce the following  4-vector $\xi_{\mu}\equiv(\frac{1}{2}\xi,-\frac{1}{2}\delta\omega_{0i}^{7})\equiv(\xi_{0},\xi_{i})$; then $\xi=2\xi_{0}$ and $\delta\omega_{0i}^{7}=-2\xi_{i}$. Hence, the relation \eqref{RelationParameters1}  takes the form
\begin{equation}
\label{RelationParameters2}
\frac{1}{2}\delta\omega_{0i}^{2} = \partial_{0}\xi_{i} + \partial_{i}\xi_{0},
\end{equation}
finally, from  the equations \eqref{deltah00}-\eqref{deltahij} and \eqref{RelationParameters2} the following gauge transformations  are identified 
\begin{equation}
\label{GT}
\delta h_{\mu\nu} = \partial_{\mu}\xi_{\nu} + \partial_{\nu}\xi_{\mu}.
\end{equation}
all these  results are in agreement with those reported in \cite{Aldair1}, thus,  our study complete and extends  those   reported in the literature.

\section{Conclusions and remarks}
In this paper a detailed $HJ$ analysis for  the higher-order  modified gravity has been performed. We introduced a new set of variables in  a different way  than  other approaches and reported in the literature, then the full set of involutive and non-involutive Hamiltonians  were identified. The correct identification of the Hamiltonians  allow us to avoid the Ostrogradsky instability by removing the terms with linear momenta, healing the canonical Hamiltonian. Furthermore, the  $HJ$ generalized brackets and  the  fundamental differential were obtained  from which  the characteristic equations and  the gauge symmetries were identified. The complete identification of the Hamiltonians allowed us to  carry out the counting of the  physical  degrees of freedom,  concluding that the modified theory and $GR$ shares the same number of physical degrees of freedom. In this manner, we have all elements to analize  the theory in the quantum context. In fact, with our perturbative $HJ$ study  either  constraints or  the generalized brackets are under control,  thus,  we could use    the tools developed in the canonical quantization of  field theories in order to make progress in this program \cite{Amorim}. Furthermore, our analysis will be relevant for the study of the theory in  the non-perturbative scenario. In fact, now  the modified theory will be full  background independent  then  we will compare the differences between the canonical  structure of $GR$  reported  in the literature \cite{Rov, Thie} and that for the  modified theory. However,  all those ideas are still in progress and will be reported soon \cite{Aldair2}.  \\
  
%----------------------------------------------------------------------------------------
Data Availability Statement: No Data associated in the manuscript\\


\begin{thebibliography}{99}
\bibitem{Einstein1} A. Einstein, The Foundation of the General Theory of Relativity, \textit{Annalen Phys} \textbf{49}, 769-822 (1916).
\bibitem{Einstein2} A. Einstein, The Field Equations of Gravitation, \textit{Sitzungsberichte}, Royal Pruss. A. of S., Berlin, 844-847 (1915).
\bibitem{Dyson} F. Dyson, A. Eddington and C. Davison, A Determination of the Deflection of Light by the Sun's Gravitational Field, from Observations Made at the Total Eclipse of May 29 1919, \textit{Phil. Trans. R. Soc. Lond A} \textbf{220}, (1920).
\bibitem{Abbott} B. Abbott \textit{et al}, Observation of Gravitational Waves from a Binary Black Hole Merger, \textit{Phys. Rev. Lett.} \textbf{116}, 061102 (2016).
\bibitem{TEHTC} The Event Horizon Telescope Collaboration, First M87 Event Horizon Telescope Results. I. The Shadow of the Supermassive Black Hole, \textit{The Astrophysical Journal Letters} \textbf{875}, 1 (2019).
\bibitem{Turyshev} S. Turyshev, Experimental Test of General Relativity: Recent Progress and Future Directions, \textit{Ups. Fiz. Nauk} \textbf{52} 1-27 (2009).
\bibitem{DeWitt} B. DeWitt, Quantum Theory of Gravity. I. The Canonical Theory, \textit{Phys. Rev.} \textbf{160}, 1113 (1967).
\bibitem{Rov} Rovelli, C. Quantum Gravity. Cambridge University Press, Cambridge (2004)
\bibitem{Thie} Thiemann, T. Modern Canonical Quantum General Relativity. Cambridge University Press, Cambridge
(2007).
\bibitem{Kiefer} C. Kiefer, Quantum Gravity, Oxford Science Publications, (2007).
\bibitem{Deser} S. Deser and P. Nieuwenhuizen, Nonrenormalizability of the Quantized Einstein-Maxwell System, \textit{Phys. Rev. Lett.} \textbf{32}, 245 (1974).
\bibitem{Weyl} H. Weyl, A New Extension of Relativity Theory, \textit{Annalen Phys.} \textbf{59}, 101-133 (1919).
\bibitem{Bach} R. Bach, On Weyl's theory of relativity and Weyl's extension of the concept of curvature tensors, \textit{Mathematische Zeitschrift} \textbf{9}, 110-135 (1921).
\bibitem{Chen} Q. Chen and Y. Ma, Hamiltonian structure and connection dynamics of Weyl gravity, \textit{Phys. Rev.} \textbf{98}, 064009 (2018).
\bibitem{Alkac} G. Alkac, M. Tek and B. Tekin, Bachian gravity in three dimensions, \textit{Phys. Rev. D} \textbf{98}, 104021 (2018).
\bibitem{Stelle} K. Stelle, Renormalization of higher-derivative quantum gravity, \textit{Phys. Rev. D} \textbf{16}, 953 (1977).
\bibitem{frad} E.S. Fradkin, A.A. Tseytlin, Renormalizable asymptotically free quantum theory of gravity, \textit{Nucl. Phys. B} \textbf{201}, 469 (1982).
\bibitem{Gib} G. W. Gibbons, Phantom Matter and the Cosmological Constant, arXiv:hep-th/0302199 (2003).
\bibitem{Wo} R.P. Woodard, Avoiding Dark Energy with 1/R Modifications of General Relativity, \textit{Lect. Notes Phys.} \textbf{720}, 403 (2007).
\bibitem{Podolsky} B. Podolsky, A Generalized Electrodynamics, \textit{Phys. Rev.} \textbf{62}, 68 (1942).
\bibitem{Pod2} Podolsky, C. Kikuchi, A Generalized Electrodynamics Part II, \textit{Phys. Rev.} \textbf{65}, 228 (1944).
\bibitem{Pod3} Podolsky, C. Kikuchi, Auxiliary Conditions and Electrostatic Interaction in Generalized Quantum Electrodynamics, \textit{Phys. Rev.} \textbf{67}, 184 (1945).
\bibitem{Pol}  A. Polyakov, Fine structure of strings, \textit{Nucl. Phys. B} \textbf{268}, 406 (1986).
\bibitem{ELi} D.A. Eliezer, R.P. Woodard, The problem of nonlocality in string theory, \textit{Nucl. Phys. B} \textbf{325}, 389 (1989).
\bibitem{Jackiw} R. Jackiw and S. Yi, Chern-Simons modification of general relativity, \textit{Phys. Rev. D} \textbf{68}, 104012 (2003).
\bibitem{Aldair1} A. Escalante and A. Pantoja, Hamiltonian analysis for higher order theories: Chern-Simons modification of general relativity, \textit{The European Physical Journal C}, under review (2022).
\bibitem{Ostrogradsky} M. Ostrogradsky, Memoires sur les equations differentielles, relatives au probleme des isoperimetres, \textit{Mem. Ac. St. Petersbourg}, 385 (1850).
\bibitem{Dirac1} P. Dirac, Generalized hamiltonian dynamics, \textit{Canadian Journal of Mathematics} \textbf{2}, 129-148 (1950).
\bibitem{Dirac2} P. Dirac, Lectures on Quantum Mechanics, Yeshiva University, New York, (1964).
\bibitem{Henneaux} M. Henneaux and C. Teitelboim, Quantization of Gauge Systems, Princeton University, (1994).  
\bibitem{Gitman1} D. Gitman, S. Lyakhovich and I. Tyutin, Hamiltonian formulation of a theory with higher derivatives, \textit{Sov. Phys. Journal} \textbf{26}, 730-734 (1983).
\bibitem{Gitman2} D. Gitman and I. Tyutin, Quantization of Fields with Constraints, Springer, (1990).
\bibitem{Barcelos} J. Barcelos and T. Dargam, Constrained analysis of topologically massive gravity, \textit{Z. Phys. C Particles and Fields} \textbf{67}, 701-705 (1995).
\bibitem{Esca} A. Escalante, Jorge Hern\'andez-Aguilar, New canonical analysis for higher order topologically massive
gravity \textit{Eur. Phys. J. C} \textbf{81}, 678, (2021).
\bibitem{Guler1} Y. Güler, Hamilton-Jacobi Theory of Discrete, Regular Constrained Systems, \textit{IL Nuovo Cimento} \textbf{100}, 267-276 (1987).
\bibitem{Guler2} Y. Güler, Hamilton-Jacobi Theory of Continuous Systems, \textit{IL Nuovo Cimento} \textbf{100}, 251-266 (1987).
\bibitem{Guler3} Y. Güler, Canonical Formulation of Singular Systems, \textit{IL Nuovo Cimento} \textbf{107}, 1389-1395 (1992).
\bibitem{Guler4} Y. Güler, Integration of Singular Systems, \textit{IL Nuovo Cimento}, \textbf{107} 1143-1149, (1992).
\bibitem{Bertin1}  M.C. Bertin, B.M. Pimentel, C.E. Valcárcel, Non-involutive constrained systems and Hamilton-Jacobi formalism, \textit{Ann. Phys.} \textbf{323}, 3137 (2008).
\bibitem{Bertin2} M.C. Bertin, B.M. Pimentel, C.E. Valcárcel, Involutive constrained systems and Hamilton-Jacobi formalism, \textit{J. Math. Phys.} \textbf{55}, 112901 (2014).
\bibitem{pimentel} N.T. Maia, B.M. Pimentel, C.E. Valcárcel, Three-dimensional background field gravity: a Hamilton-Jacobi analysis, \textit{Class. Quantum Grav.} \textbf{32}, 185013 (2015).
\bibitem{esca} A. Escalante, A. Pantoja, The Hamilton-Jacobi analysis and covariant description for three-dimensional Palatini theory plus a Chern-Simons term, \textit{Eur. Phys. J. Plus} \textbf{134}, 437 (2019).
\bibitem{esca2} A. Escalante, M. Eduardo Hernández-García, The Hamilton-Jacobi characteristic equations for three-dimensional Ashtekar gravity, \textit{Eur. Phys. J. Plus} \textbf{135}, 245 (2020).
\bibitem{plus} R. P. Woodard, The Theorem of Ostrodradsky, arXiv:1506.02210v2.
\bibitem{Ganz} A. Ganz and K. Noui, Reconsidering the Ostrogradsky theorem: higher-derivatives Lagrangians, ghost and degeneracy, \textit{Class. Quantum Grav.} \textbf{38}, 075005 (2021).
\bibitem{Tai} Tai-jun Chen, M. Fasiello, Eugene A. Lim, Andrew J. Tolley, Higher derivative theories with constraints: Exorcising Ostrogradski's Ghost, 	\textit{JCAP} \textbf{130}, 042, (2013).
%Repetida en [19] \bibitem{Woo} R. P. Woodard, \textit{Avoiding Dark Energy with 1/R Modifications of Gravity}, Lect.Notes Phys.720:403-433, (2007).
\bibitem{Frankel} T. Frankel, \textit{The Geometry of Physics} 3rd, Cambridge University Press, (2012).
\bibitem{Furi} H. Fuhri, S. Hortner, Phys. Rev. D 103, 105014, (2021).
\bibitem{Bertin} M. Bertin, B. Pimentel, C. Valcarcel and G. Zambrano, Hamilton-Jacobi formalism for linearized gravity, \textit{Class. Quantum Grav.} \textbf{28}, 175015 (2011).
\bibitem{Ber} M.C. Bertin, B.M. Pimentel, C.E. Valcárcel, G.E.R. Zambrano, Involutive constrained systems and Hamilton-Jacobi formalism, \textit{J. Math. Phys.} \textbf{55}, 112901 (2014).
\bibitem{Amorim} R. Amorim and J. Barcelos, Functional versus canonical quantization of nonlocal massive vector-gauge theory, \textit{J. Math. Phys.} \textbf{40}, 585 (1999).
\bibitem{Aldair2} A. Escalante and A. Pantoja, \textit{The perturbative and non-perturbative canonical analysis of the Chern-Simons modification of General Relativity}, in progress.
%\bibitem{Green2} A. Green, On Generalizing Boson Field Theories, \textit{Phys. Rev.} \textbf{75}, 1926 (1949).
%\bibitem{Banerjee} R. Banerjee, P. Mukherjee, B. Paul, Gauge symmetry and W-algebra in higher derivative systems, \textit{Journal of High Energy Physics}, 85 (2011).
%\bibitem{Escalante1} A. Escalante and O. Rodriguez, Hamiltonian dynamics and gauge symmetry for three-dimensional Palatini theory with cosmological constant, \textit{Journal of High Energy Physics}, 73 (2014).
%\bibitem{Escalante2} A. Escalante and M. Rodriguez, Canonical and symplectic analysis of actions describing linearized gravity, \textit{Eur. Phys. J Plus} \textbf{134}, 152 (2019).
%\bibitem{Faddeev} L. Faddeev and R. Jackiw, Hamiltonian Reduction of Unconstrained and Constrained Systems, \textit{Phys. Rev. Lett.} \textbf{60}, 1962 (1988).
\end{thebibliography}
\end{document}